**Black Arsenic-Phosphorus: Layered Anisotropic Infrared Semiconductors with Highly Tunable Compositions and Properties**

*Bilu Liu1, Marianne Köpf2, Ahmad A. Abbas1, Xiaomu Wang3, Qiushi Guo3, Yichen Jia3, Fengnian Xia3, Richard Weihrich4, Frederik Bachhuber4, Florian Pielnhofer4, Han Wang1, Rohan Dhall1, Stephen B. Cronin1, Mingyuan Ge1, Xin Fang1, Tom Nilges2, Chongwu Zhou1\**

Dr. B. Liu, A. Abbas, Prof. H. Wang, R. Dhall, Prof. S. B. Cronin, M. Ge, X. Fang, Prof. C. Zhou
Ming Hsieh Department of Electrical Engineering, University of Southern California, Los Angeles, California, 90089, USA
E-mail: chongwuz@usc.edu

M. Köpf , Prof. T. Nilges
Technische Universität München, Department of Chemistry, Lichtenbergstraße 4, Garching b. München 485748, Germany

Dr. X. Wang, Q. Guo, Y. Jia, Prof. F. Xia
Department of Electrical Engineering, Yale University, New Haven, Connecticut 06511, USA

Prof. R. Weihrich, F. Bachhuber, F. Pielnhofer
Institut für Anorganische Chemie, Universität Regensburg, Universitätsstraße 31, Regensburg 93040, Germany

**Keywords:** Black arsenic, black phosphorus, layered materials, anisotropic, infrared semiconductor, band gap

*Main text*

In recent years, two-dimensional (2D) layered materials, including graphene,[1-3] hexagonal boron nitride (h-BN),[4-6] transition metal dichalcogenides (TMDCs),[4, 5, 7-11] and black phosphorus (b-P),[12-16] have attracted significant interest due to their unique electronic, optical, and mechanical properties. To date, these 2D layered materials have covered a wide electromagnetic spectral range including metals, semimetals, semiconductors, and insulators.

Specifically, monolayer graphene is a semimetal which exhibits a zero band gap in its natural state, and strong electrical field can be used in order to create a band gap in bilayer graphene.[17] On the other end of the spectrum, h-BN is an insulator with a large band gap of around 6 eV, which extends into the middle ultraviolent regime. TMDCs have a formula of $MX_2$ where M is a transition metal element and X is a chalcogen element. TMDCs can be either metals or semiconductors, depending on the transition elements.[18] The most heavily studied TMDCs so far (M=Mo or W, and X=S, Se, or Te) possess band gaps in the range from around 1.0 eV to 1.9 eV via tuning the chemical compositions, layer numbers, and strains of the materials.[8, 18-22] These TMDC materials fill up part of the gap between graphene and h-BN in the electromagnetic spectra. One of the newest members of the 2D layered material family is b-P,[12-14] which is a semiconductor exhibiting a moderate band gap of around 0.3 eV in its bulk form and up to 2.2 eV as in a monolayer,[23] further pushing the band gap of layered materials to the middle-wavelength infrared regime (MWIR, 0.3 eV corresponds to a wavelength of 4.13 μm). Therefore, a broad spectral range has now been covered by the above-mentioned layered materials. Nevertheless, there is a technologically-important wavelength range, the long-wavelength infrared (LWIR) regime, which still cannot be readily covered by the 2D layered materials mentioned above. This particular electromagnetic spectral range is important for a range of applications such as range finding using LIDAR (light radar) systems, because the earth atmosphere (mostly $H_2O$, $O_3$, and $CO_2$) has good transparency in this spectral range (starting at around 8 μm). Therefore, new layered semiconducting materials that cover this distinct electromagnetic spectral range are desirable for the realization of devices and systems based on all 2D layered materials.

Alloying is a general strategy that has been used for tuning the properties of materials for thousands of years. Graphene, BN, and TMDCs with tunable properties can be synthesized via such alloying or doping methods.[6, 19, 21, 22] It has been shown two decades ago that arsenic can be incorporated into b-P via a high pressure process, and these materials show superconducting properties at around 10 K.[24] Here, we introduce a family of layered semiconductors, black arsenic-phosphorus (b-AsP), via a new synthetic approach adopting the alloying strategy. Our method involves the synthesis of layered b-AsP materials with different and tunable compositions (b-As$_x$P$_{1-x}$, with x in the range of 0 to 0.83), using a novel mineralizer-assisted short way chemical transport reaction (see Methods). More importantly, thanks to the good tunability of the chemical compositions, we demonstrate that these layered b-AsP materials have fully tunable band gaps and optical properties that can cover long wavelengths down to around 0.15 eV (corresponds to a wavelength of 8.27 μm, LWIR regime). A new synthetic route is used for the synthesis of b-AsP in this study, which uses ultrapure grey arsenic and red phosphorus as starting materials. This new route was developed based on our earlier work[25, 26] and inspired by the work on the synthesis of b-P.[16, 27, 28] The as-synthesized materials are black shin flakes or needle-like structures. X-ray diffraction (Supplementary Figure S1) [26] and transmission electron microscopy (Supplementary Figure S2) characterization shows that this family of layered b-AsP materials, similar to b-P, have orthorhombic structure with a puckered honeycomb lattice (A17 type structure, see Figures 1a and 1b). Arsenic and phosphorus atoms are distributed in the materials, and the volume of the unit cell increases with increasing arsenic content (Supplementary Figure S3). Therefore, these b-AsP materials have strong in-plane covalent bonding and weak interlayer van der Walls interactions, sharing the common features with many other layered 2D materials.[4, 7, 8, 11-14, 18, 23, 29]

These layered b-AsP materials can be mechanically-exfoliated into thin flakes down to atomic layers. Figure 1c shows an atomic-force microscopy (AFM) image of a b-As$_{0.83}$P$_{0.17}$ flake with a thickness of 1.3 nm, corresponding to two atomic layers. Interestingly, during exfoliation experiments, we found that the difficulty in exfoliating these materials increases as the arsenic content increases in the b-AsP materials; an indication of the enhanced inter-layer interaction for materials with high arsenic content. This phenomenon may shed light on future studies of the strength of interlayer van der Walls interactions in layered materials of group V elements.

To study the charge transport properties in these materials, we fabricated back gate field-effect transistors (FETs) using exfoliated b-As$_{0.83}$P$_{0.17}$ flakes as channel materials. The samples were exfoliated onto Si substrates covered with 300 nm SiO$_2$ layer and the devices were fabricated using e-beam lithography. The metal contact was Ti/Au with thickness of 1 nm/50 nm or 5 nm/50 nm. Figure 2a is a three dimensional AFM image of a typical b-AsP FET possesses a channel length of 1.3 μm and a channel width of 2 μm. We found that for thick b-AsP samples (>20-30 nm in thickness), the devices are highly conductive and exhibit very weak gate dependence due to screening effect (see Supplementary Figure S4). When the flakes become thin, increased gate modulation to the channel conductance was observed. Figure 2b shows the transfer curves of a 15-nm-thick b-As$_{0.83}$P$_{0.17}$ flake at different bias voltages. This device shows a hole mobility of 110 cm$^2$/V·s under two-terminal configuration. Figure 2c shows transfer curve of a thin (5 nm) b-As$_{0.83}$P$_{0.17}$ flake. This device shows ambipolar transport behavior and an on/off current ratio of around 2000. These transport measurements reveal the semiconducting nature of b-AsP materials.

One major motivation of studying layered b-AsP materials is whether we can tune the band gaps of these materials by tuning the chemical compositions. We used polarization-resolved infrared absorption spectroscopy to measure the band gaps as well as anisotropic behavior of b-AsP. Figure 3a shows infrared absorption spectra of b-As$_{0.83}$P$_{0.17}$, which has the highest arsenic content in this study. The results show that this material has an absorption edge of around 1250 cm$^{-1}$, corresponding to a band gap of 0.15 eV. Noticeably, this band gap is only half of the band gap of b-P (around 0.3 eV) and reaches the LWIR regime. Moreover, the polarization-resolved infrared absorption spectra clearly show the anisotropic absorption behavior of b-As$_{0.83}$P$_{0.17}$ (Figure 3a). A full list of polarization-resolved spectra with an angle step of 15 degree are shown in Supplementary Information Figure S5, and for clarity, only four spectra are shown in Figure 3a. Quantitative analysis of the absorption intensity at 2200 cm$^{-1}$ is exhibited in Figure 3b. The absorption maximum is found at an angle of 135 degree, which is close to the x-axis of the flake (see Figure 1a and 1b) with an error of less than 7.5 degree (due to the use of 15 degree angle step during experiments).[14] Accordingly, the absorption minimum is located at an angle of 45 degree, with is close to the y-axis of the flake.

We performed systematic polarization-resolved infrared absorption studies on a family of b-As$_x$P$_{1-x}$ with different x values of 0, 0.25, 0.4, and 0.83. Figure 3c shows a collection of infrared absorption spectra for b-AsP with different compositions. A clear shift of the absorption edges to shorter wavelengths with increasing x in b-As$_x$P$_{1-x}$ is observed, confirming that the band gaps of b-As$_x$P$_{1-x}$ change with the chemical composition of the material. This result provides the basis for applications of these layered b-AsP materials in a variety of fields including electronics and optoelectronics, such as middle to long wavelength infrared photo-detection and optical imaging.

We have conducted infrared absorption measurements on these materials, and collected a total of around 80 spectra for statistical analysis. Figure 3d shows a summary of the results based on multiple spectra taken for each composition. It should be noted that all the flakes measured in Figure 3d are relatively thick ones with thicknesses greater than 30 nm. One can clearly see that the band gaps decrease with increasing arsenic content in the b-As$_x$P$_{1-x}$ materials. Noticeably, the band gaps of these layered b-As$_x$P$_{1-x}$ materials can be fully tuned over the range from 0.3 eV (in the case of b-As$_0$P$_1$, i.e., pure b-P) to 0.15 eV (in b-As$_{0.83}$P$_{0.17}$). Further increasing the arsenic content, for example, up to pure black arsenic (b-As, here we refer to orthorhombic structure arsenic, A17 type structure), may push the band gap to the even longer wavelength regime. Recent revPBE-van der Waals calculations show that pure b-As is a metal with zero band gap in bulk and exhibits high charge carrier mobilities in few layer form.[29] So far, there has been no success in the synthesis of pure b-As due to its metastability and the lack of suitable synthesis strategies.[26] We note that due to the applied short way vapor transport reaction, a variation of the composition within one sample is possible. We have crosschecked multiple crystals of each batch by energy-dispersive X-ray spectroscopy (EDX) and found that the contents of arsenic and phosphorus (As/P ratio) fall into a reasonably narrow range. Therefore, a small variation of composition of the materials does not change the big picture of composition-dependent band gaps we reported in this study.

Interestingly, one can discern from Figure 3d that the band gaps of the b-$As_xP_{1-x}$ decrease sharply when the arsenic content changed from 0 to 0.25, and then decrease less significantly when the arsenic content changes from 0.25 to 0.4 and further to 0.83. The results suggest that the band gaps do not scale linearly with the content of arsenic or phosphorus. The general trend of composition-dependent band gaps obtained from infrared absorption measurements shows agreement with our density-functional theory (DFT) calculations shown in Supplementary Table S1.

We have further substantiated the anisotropic properties of these b-$As_xP_{1-x}$ materials via polarization-resolved Raman spectroscopy studies. Here, we use b-$As_{0.83}P_{0.17}$ as an example to present this behavior. In our experiments, light is linearly incident from the z-direction, is polarized in the x-y plane, and the detection is in the z direction and unpolarized. The wavelength of laser is 532 nm and the spot size is ~1 μm. We first compared the Raman spectra of b-$As_{0.83}P_{0.17}$ with pure b-P and b-As. For b-As, since there have been no reports on the synthesis and Raman characterization of this material so far, we therefore calculated its Raman spectra. It can be seen that the characteristic peaks of these systems are quite different (Supplementary Figure S6, Figure S7, and Table S2), indicating different vibrational properties. The Raman spectra of b-P have been well documented, and three major peaks are observed, including the out-of-plane $A_g^1$ peak at 363 cm$^{-1}$, the in-plane $B_{2g}$ peak at 440 cm$^{-1}$, and the in-plane $A_g^2$ peak at 467 cm$^{-1}$. These experimental observations are consistent with our calculations (Supplementary Figure S7 and Table S2) and recent reports on the Raman spectra of b-P.[13, 14, 23] For b-As, our calculations indicate the $A_g^1$ peak at 237 cm$^{-1}$, the $B_{2g}$ peak at 241 cm$^{-1}$, and the $A_g^2$ peak at 273 cm$^{-1}$ (Supplementary Figure S7 and Table S2).

As a contrast, b-As$_{0.83}$P$_{0.17}$ exhibits more Raman peaks (Figure 4a and Supplementary Figure S5) than both b-P and b-As, in good accordance with the occurrence of heteroatomic arsenic-phosphorus bonds. One can see that the Raman spectra of b-As$_{0.83}$P$_{0.17}$ can be divided into three major regimes, low frequency (~200-300 cm$^{-1}$), medium frequency (~300-380 cm$^{-1}$), and high frequency (~380-500 cm$^{-1}$) regimes. Dependent on the grade of substitution of phosphorus by arsenic, we observed that the relative intensity of peaks at each regime varies significantly. Specifically for b-As$_{0.83}$P$_{0.17}$, the three peaks from low frequency regime (~200-300 cm$^{-1}$), i.e., 224, 233, and 256 cm$^{-1}$, dominate the spectra. We have assigned these peaks to $A_g^1$, $B_{2g}$, and $A_g^2$ modes of b-As$_{0.83}$P$_{0.17}$. There are systematical blue shifts in peak positions compared with our calculated data on pure b-As (Table S2), because the existence of phosphorus atoms in b-As$_{0.83}$P$_{0.17}$ can blue shift the Raman peaks of b-As. With increasing the phosphorus content, the peaks gradually shift toward medium and high frequency regimes (Figure 4a).

In the b-AsP materials, several different bonds may exist including arsenic-arsenic, phosphorus-phosphorus, and arsenic-phosphorus. Therefore, it is reasonable that b-AsP would exhibit more Raman peaks than pure b-P or b-As, as Raman scattering probes a range corresponding to second or further neighbors and is sensitive to the short-range order beyond the nearest neighbors. In general, we expect a defined separation of the homoatomic arsenic-arsenic and phosphorus-phosphorus peaks and the heteroatomic arsenic-phosphorus peaks in the spectra. The homoatomic phosphorus-phosphorus peaks in b-P are located in the spectral region larger than 380 cm$^{-1}$. Homoatomic arsenic-arsenic peaks can be found in the region below 300 cm$^{-1}$. Raman studies on b-As$_x$P$_{1-x}$ flakes with different x values reveal that for materials with high content of phosphorus, the characteristic peaks from b-P largely remain similar, except a slight red shift of the b-P peaks (Figure 4a). This finding is in accordance with the increasing amount of phosphorus-phosphorus bonds and the decreasing contribution of arsenic-arsenic bonds, as well as the change in bond strength and bond lengths in the system upon substitution. With increasing arsenic concentration, the phosphorus-phosphorus peaks gradually become weaker and weaker, and new arsenic-phosphorus peaks appear in the middle frequency regime and arsenic-arsenic peaks appear in the low frequency regime (Figure 4a). We also calculated the Raman spectra of b-AsP systems and found that the phosphorus-arsenic peaks are predicted to occur in the range from 300 to 380 $cm^{-1}$ (Supplementary Figure S7 and Table S2), showing good agreement with the experimental observations.

Now let us focus on the polarization behavior of these Raman peaks. Figure 4b presents the polarization-resolved Raman spectra of a b-$As_{0.83}P_{0.17}$ flake at different incident laser polarization angles. A clear polarization-dependence of the Raman behavior is observed, further confirming the anisotropic properties of these b-AsP materials as revealed from infrared absorption studies. The three major peaks (located at 256, 233, and 224 $cm^{-1}$) show polarization-dependent Raman intensity, as exhibited in Figures 4c. The observed angle-dependent Raman intensities are related to the structure of Raman tensors for each peak. Via these polarization-resolved Raman experiments, we found the following (see Supplementary Figure S6 and main text Figure 4c): i) The exfoliated thin b-AsP flakes (<5 nm) are still crystalline since they exhibit similar Raman features with the bulk, ii) Raman spectra from thick (>20 nm) and thin (<5 nm) flakes show similar characteristics in terms of the number of peaks, peak frequency, and peak-peak distance, iii) There is no noticeable shift in the peak frequency for spectra collected at different incident laser polarization angles, and iv) The maximum intensities for different peaks appear at different polarization angles, which is an indication of significant differences in their Raman tensors. We have performed polarization-resolved Raman studies on seven flakes, and they show consistent polarization-dependent Raman intensity phenomena.

**Conclusion**

In conclusion, we have introduced a new family of layered semiconducting materials, black arsenic-phosphorus (b-$As_xP_{1-x}$). We show that these layered materials can be rationally synthesized with highly tunable chemical compositions, and can be exfoliated down to a few atomic layers. Electron transport measurements reveal the semiconducting nature of these materials. Through infrared absorption studies, we demonstrate that these layered b-AsP materials are semiconductors with tunable electronic and optical properties, via tuning the chemical compositions during material synthesis. Noticeably, this family of b-AsP materials fills up an interesting and technologically-important gap in the LWIR regime in the electromagnetic spectra, which cannot be readily achieved by other 2D layered materials so far. Polarization-resolved infrared absorption and polarization-resolved Raman scattering studies reveal the in-plane anisotropic properties of these materials. Therefore, the layered b-AsP materials reported in this study contribute to the fabrication of functional devices based on all 2D layered materials. We envision that this family of new layered semiconductors with tunable band gaps in the LWIR regime, as well as with anisotropic properties, may find unique applications for electronic and optoelectronics devices operate at infrared regime.

**Methods**

*Synthesis of composition-tunable black arsenic phosphorus (b-$As_xP_{1-x}$).* b-$As_xP_{1-x}$ samples with different nominal compositions were synthesized using a vapor transport method. Specifically, b-P (b-$As_0P_1$), b-$As_{0.25}P_{0.75}$, and b-$As_{0.4}P_{0.6}$ samples were prepared from a mixture of grey arsenic (Chempur, 99.9999%) and red phosphorus (Chempur, 99.999$^+$ %) with molar ratio of 0:1, 1:3, and 2:3, respectively. Tin/tin (IV) iodide (Sn/$SnI_4$ = 10/5 mg per 250 mg batch) were added to the starting materials and acted as mineralizer additives to allow phase formation and crystal growth. All chemicals were enclosed in evacuated silica glass ampoules (length: 100 mm, inner diameter: 8 mm) during the reaction.

Synthesis of b-P and b-$As_{0.25}P_{0.75}$ samples was performed in a Nabertherm furnace (L3/11/P330). The following heating program was applied. i) Heating the starting materials up to 650 °C within 8 hours and holding for 5 hours at this temperature, iii) Cooling down to 550 °C within 7.5 hours and holding for 6 hours at this temperature, iii) Further cooling down to 500 °C and holding for 8 hours at this temperature, and IV) Cooling down to room temperature within 20 hours. The b-$As_{0.4}P_{0.6}$ samples were synthesized using a Nabertherm furnace (L3/11/P320) and the heating program was changed to: i) Heating to 550 °C within 8 hours and holding for 6 hours at this temperature, ii) Cooling down to 500 °C within 2 hours and holding for 8 hours at this temperature, and iii) Cooling down to room temperature within 20 hours.

The b-$As_{0.83}P_{0.17}$ samples were prepared from a mixture of gray arsenic (Chempur, 99.9999%) and red phosphorus (Chempur, 99.999$^+$ %) and with a molar ratio 83:17. The mineralizing agent was lead (II) iodide ($PbI_2$, 12 mg per 625 mg batch). The chemicals were enclosed in evacuated silica glass ampoules during reaction (length: 100 mm, inner diameter: 10 mm). Synthesis was performed in a Nabertherm furnace (L3/11/P330). The following heating program was applied, i) Heating up to 550 °C within 8 hours and holding for 84 hours at this temperature, and ii) Cooling down to room temperature within 20 hours.

After synthesis, bulk b-AsP samples were mechanically-exfoliated into flakes using a Scotch tape for optical and electrical measurements.

***b-AsP characterization.*** The AFM studies were performed using a Digital Instrument Dimensional 3100 AFM in tapping mode. We found that thin b-AsP flakes, similar to b-P samples, experience degradation when exposed to air. For b-AsP flakes covered with PMMA layer or stored in argon glove box, they were stable and no noticeable degradation was observed over several month periods (Supplementary Figure S8).

*CAUTION:* In the case of handling b-AsP in oxygen or humid atmosphere, the formation of $As_2O_3$, which represents the major toxic component in the system, cannot be fully suppressed. Consequently, cautions should be taken when handling this material.

***Device fabrication and measurements.*** Back-gate b-AsP FETs were fabricated using e-beam lithography. The metal contact was Ti/Au with thicknesses of 1/50 nm or 5/50 nm. The device measurements were performed using Agilent 4156B semiconductor parameter analyzer under two-terminal configuration.

***Polarized infrared absorption studies.*** Polarization-resolved infrared spectroscopic studies were performed in the 600–6,000 $cm^{-1}$ range using a Bruker optics Fourier transfer infrared spectrometer (Vertex 70) integrated with a Hyperion 2000 microscope system. The polarization angle of the incident light was adjusted using an infrared polarizer with an angle step of 15 or 30 degree. A 36x objective was used to collect the signal and the beam size was around 35 μm. The experiments were done in ambient condition.

***Polarized Raman studies.*** Polarization-resolved Raman studies were performed using a Renishaw micro Raman system with an excitation laser wavelength of 532 nm and beam size of ~1 μm. The laser incident from z-direction and was linearly polarized in the x-y plane. The polarization direction of the laser was adjusted using a half-wavelength plate with an angle step of 15 degree. The detection was along the z direction and was un-polarized. The samples were loaded into a vacuum chamber, and the laser power was kept around 5 μW to avoid sample degradation during Raman measurements. The spectrum integration time was 60 seconds. No obvious degradation of b-AsP was observed after a series of Raman measurements in vacuum.

*DFT calculations.* First principles calculations were performed within the framework of DFT with GGA-PBE functionals and Grimme D2 corrections for van-der-Waals interactions, to calculate the band gap and vibrational properties of b-AsP system. More details of the calculations are shown in the Supplementary Information.

**Associated Content**

*Acknowledgements* We acknowledge Feng Wang and Jonghwan Kim for helpful discussions. The work at University of Southern California was supported by the Office of Naval Research (ONR) and the Air Force Office of Scientific Research (AFOSR). We would like to acknowledge the collaboration of this research with King Abdul-Aziz City for Science and Technology (KACST) via The Center of Excellence for Nanotechnologies (CEGN). The work at Technische Universität München and Universität Regensburg was supported by the DFG within the priority research program 1415. M.K. would like to thank the TUM Graduate School for support.

*Author contributions*

B.L. and C.Z. conceived the idea. B.L., C.Z, and T.N. initiated the study. M.K. prepared black arsenic-phosphorus samples under supervision of T.N. X.X., Q.G., J.J., F.X., H.W., and B.L. performed polarization-resolved IR absorption measurements and analysis. B.L., R.D., and S.C. performed polarization-resolved Raman studies and analysis. R.W., F. B., and F.P. performed the DFT simulations. B.L. and A.A. fabricated devices and did the measurements. M.G. involved in discussion and schematic drawing. X.F. conducted the EDX measurements. B.L., M.K., T.N., C.Z. and R.W. wrote the manuscript with feedback from all others.

*References*

[1] K. S. Novoselov, A. K. Geim, S. V. Morozov, D. Jiang, Y. Zhang, S. V. Dubonos, I. V. Grigorieva, A. A. Firsov, *Science* **2004,** *306*, 666.

[2] K. S. Novoselov, A. K. Geim, S. V. Morozov, D. Jiang, M. I. Katsnelson, I. V. Grigorieva, S. V. Dubonos, A. A. Firsov, *Nature* **2005,** *438*, 197.
[3] Y. B. Zhang, Y. W. Tan, H. L. Stormer, P. Kim, *Nature* **2005,** *438*, 201.
[4] K. S. Novoselov, D. Jiang, F. Schedin, T. J. Booth, V. V. Khotkevich, S. V. Morozov, A. K. Geim, *P Natl Acad Sci USA* **2005,** *102*, 10451.
[5] J. N. Coleman, M. Lotya, A. O'Neill, S. D. Bergin, P. J. King, U. Khan, K. Young, A. Gaucher, S. De, R. J. Smith, I. V. Shvets, S. K. Arora, G. Stanton, H. Y. Kim, K. Lee, G. T. Kim, G. S. Duesberg, T. Hallam, J. J. Boland, J. J. Wang, J. F. Donegan, J. C. Grunlan, G. Moriarty, A. Shmeliov, R. J. Nicholls, J. M. Perkins, E. M. Grieveson, K. Theuwissen, D. W. McComb, P. D. Nellist, V. Nicolosi, *Science* **2011,** *331*, 568.
[6] Z. Liu, L. Ma, G. Shi, W. Zhou, Y. Gong, S. Lei, X. Yang, J. Zhang, J. Yu, K. P. Hackenberg, A. Babakhani, J. C. Idrobo, R. Vajtai, J. Lou, P. M. Ajayan, *Nat Nanotechnol* **2013,** *8*, 119.
[7] B. Radisavljevic, A. Radenovic, J. Brivio, V. Giacometti, A. Kis, *Nat Nanotechnol* **2011,** *6*, 147.
[8] K. F. Mak, C. Lee, J. Hone, J. Shan, T. F. Heinz, *Phys Rev Lett* **2010,** *105*, 136805.
[9] Y. Zhang, T. R. Chang, B. Zhou, Y. T. Cui, H. Yan, Z. K. Liu, F. Schmitt, J. Lee, R. Moore, Y. L. Chen, H. Lin, H. T. Jeng, S. K. Mo, Z. Hussain, A. Bansil, Z. X. Shen, *Nat Nanotechnol* **2014,** *9*, 111.
[10] S. Najmaei, Z. Liu, W. Zhou, X. L. Zou, G. Shi, S. D. Lei, B. I. Yakobson, J. C. Idrobo, P. M. Ajayan, J. Lou, *Nat Mater* **2013,** *12*, 754.
[11] A. Splendiani, L. Sun, Y. B. Zhang, T. S. Li, J. Kim, C. Y. Chim, G. Galli, F. Wang, *Nano Lett* **2010,** *10*, 1271.
[12] L. Li, Y. Yu, G. J. Ye, Q. Ge, X. Ou, H. Wu, D. Feng, X. H. Chen, Y. Zhang, *Nat Nanotechnol* **2014,** *9*, 372.
[13] H. Liu, A. T. Neal, Z. Zhu, Z. Luo, X. F. Xu, D. Tomanek, P. D. D. Ye, *Acs Nano* **2014,** *8*, 4033.
[14] F. Xia, H. Wang, Y. Jia, *Nat Commun* **2014,** *5*, 4458.
[15] S. P. Koenig, R. A. Doganov, H. Schmidt, A. H. C. Neto, B. Ozyilmaz, *Appl Phys Lett* **2014,** *104*.
[16] M. Köpf, N. Eckstein, D. Pfister, C. Grotz, I. Krüger, M. Greiwe, T. Hansen, H. Kohlmann, T. Nilges, *Journal of Crystal Growth* **2014,** *405*, 6.
[17] Y. Zhang, T. T. Tang, C. Girit, Z. Hao, M. C. Martin, A. Zettl, M. F. Crommie, Y. R. Shen, F. Wang, *Nature* **2009,** *459*, 820.
[18] Q. H. Wang, K. Kalantar-Zadeh, A. Kis, J. N. Coleman, M. S. Strano, *Nat Nanotechnol* **2012,** *7*, 699.
[19] Y. Gong, Z. Liu, A. R. Lupini, G. Shi, J. Lin, S. Najmaei, Z. Lin, A. L. Elias, A. Berkdemir, G. You, H. Terrones, M. Terrones, R. Vajtai, S. T. Pantelides, S. J. Pennycook, J. Lou, W. Zhou, P. M. Ajayan, *Nano Lett* **2014,** *14*, 442.
[20] S. B. Desai, G. Seol, J. S. Kang, H. Fang, C. Battaglia, R. Kapadia, J. W. Ager, J. Guo, A. Javey, *Nano Lett* **2014,** *14*, 4592.
[21] Q. Feng, Y. Zhu, J. Hong, M. Zhang, W. Duan, N. Mao, J. Wu, H. Xu, F. Dong, F. Lin, C. Jin, C. Wang, J. Zhang, L. Xie, *Adv Mater* **2014,** *26*, 2648.
[22] H. L. Li, X. D. Duan, X. P. Wu, X. J. Zhuang, H. Zhou, Q. L. Zhang, X. L. Zhu, W. Hu, P. Y. Ren, P. F. Guo, L. Ma, X. P. Fan, X. X. Wang, J. Y. Xu, A. L. Pan, X. F. Duan, *J Am Chem Soc* **2014,** *136*, 3756.
[23] X. M. Wang, A. M. Jones, K. L. Seyler, V. Tran, Y. C. Jia, H. Zhao, H. Wang, L. Yang, X. D. Xu, F. N. Xia, *arXiv:1411.1695v1* **2014**.
[24] I. Shirotani, J. Mikami, T. Adachi, Y. Katayama, K. Tsuji, H. Kawamura, O. Shimomura, T. Nakajima, *Phys Rev B* **1994,** *50*, 16274.
[25] S. Lange, M. Bawohl, R. Weihrich, T. Nilges, *Angew Chem Int Ed* **2008,** *47*, 5654.

[26] O. Osters, T. Nilges, F. Bachhuber, F. Pielnhofer, R. Weihrich, M. Schoneich, P. Schmidt, *Angew Chem Int Ed* **2012,** *51*, 2994.
[27] T. Nilges, M. Kersting, T. Pfeifer, *Journal of Solid State Chemistry* **2008,** *181*, 1707.
[28] S. Lange, P. Schmidt, T. Nilges, *Inorg Chem* **2007,** *46*, 4028.
[29] Z. Y. Zhang, D. Z. Yang, Y. H. Wang, D. S. Xue, M. S. Si, G. P. Zhang, *Arxiv: 1411.3165v1* **2014**.

**Figure Legends.**

**Figure 1. Structure and exfoliation of layered black arsenic-phosphorus (b-AsP).** (a) Side view and (b) Top view of b-AsP. This family of layered materials have orthorhombic lattice with puckered honeycomb structure (A17 type). The arsenic and phosphorus atoms are distributed within each layer and their concentrations are tunable. (c) An AFM image of an exfoliated bilayer b-$As_{0.83}P_{0.17}$ flake with a thickness of ~1.3 nm.

**Figure 2. FET of back-gate black arsenic-phosphorus.** (a) An AFM image of a b-AsP FET. The channel length and width of the device are 1.3 and 2 μm as highlighted using white arrows. (b) Transfer curves of a representative thick b-$As_{0.83}P_{0.17}$ flake. The thickness of the flake is around 15 nm, and the hole mobility of this device is 110 $cm^2$/V·s. (c) Transfer curve of a thin b-$As_{0.83}P_{0.17}$ flake in semi-log scale and linear scale (inset). The device shows am-bipolar behavior and an on/off current ratio of 1.9 x $10^3$ at hole side (p-branch). The thickness of the flake is around 5 nm. The substrates are Si/$SiO_2$ (300 nm).

**Figure 3. Anisotropic layered black arsenic-phosphorus semiconductors with tunable electronic and optical properties.** (a) Polarization-resolved infrared absorption spectra of b-$As_{0.83}P_{0.17}$ materials at different polarization angles, showing clearly polarization-dependent absorption behavior. The inset is the optical microscopy image of the flake and the polarization angle indication. The thickness of this flake is 122 nm based on AFM measurements. For clarify, only four curves with angles of 45, 75, 105, and 135 degrees were shown. A collection of all 13 spectra from 0 to 165 degrees are shown in Supplementary Figure S5. (b) Dependence of the extinction intensity at 2200 $cm^{-1}$ with polarization angles. (c) Plots of infrared absorption of different b-$As_xP_{1-x}$ samples, showing the involution of band gaps with tuning x in b-$As_xP_{1-x}$ materials. The x is the nominal composition of arsenic, which corresponds to 0, 0.25, 0.40, and 0.83 in this work. Note that the absolute extinction values of the spectra in plot c cannot be compared directly, since the flake thicknesses and sizes are different for different samples. (d) Summary of x-dependent band gaps of b-$As_xP_{1-x}$. Each data point corresponds to a measurement from either different polarization angle of the same flake, or from different flakes of the same composition. The thickness of the b-AsP flakes are >30 nm in these IR measurements.

**Figure 4. Polarization-resolved micro-Raman spectroscopic characterization of b-$As_{0.83}P_{0.17}$.** (a) Raman spectra of b-AsP with different chemical compositions, showing the evolution of Raman peaks with changing arsenic content. (b) Polarization-dependent Raman spectra of a b-$As_{0.83}P_{0.17}$ sample at different polarization angles of 0, $30^0$, $60^0$, and $90^0$. Clear polarization-dependent Raman intensities can be discerned. The inset is the optical microscopy image of the flake and the polarization angle indication. (c) Polar maps of the intensity of Raman peaks at 256, 233, and 224 cm$^{-1}$, which corresponds to $A_g^1$, $B_{2g}$, and $A_g^1$ modes of b-$As_{0.83}P_{0.17}$, showing clear polarization-dependent Raman intensities. The excitation laser wavelength is 532 nm and the laser size is around 1 μm.

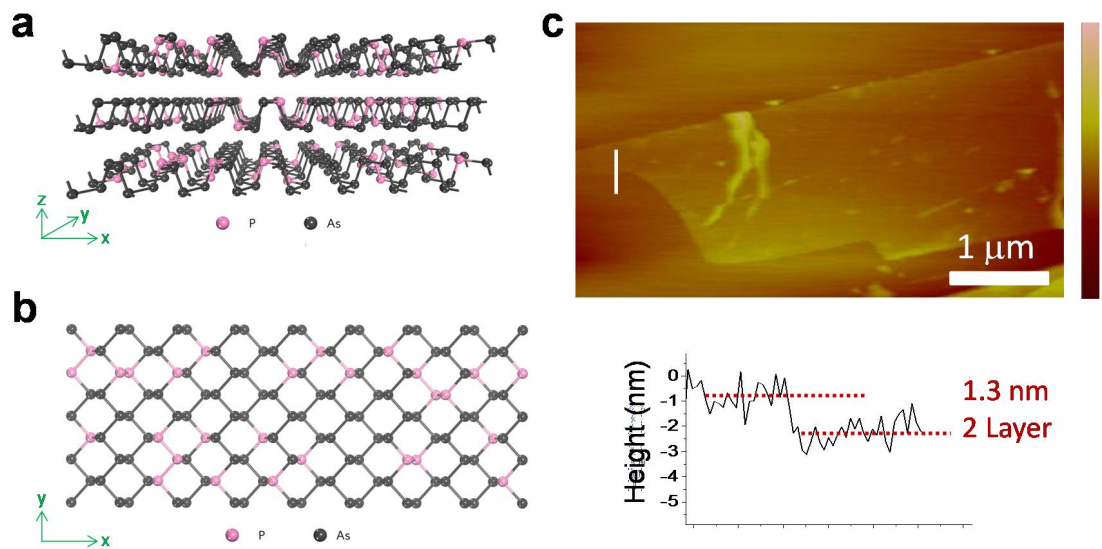

**Figure 1**

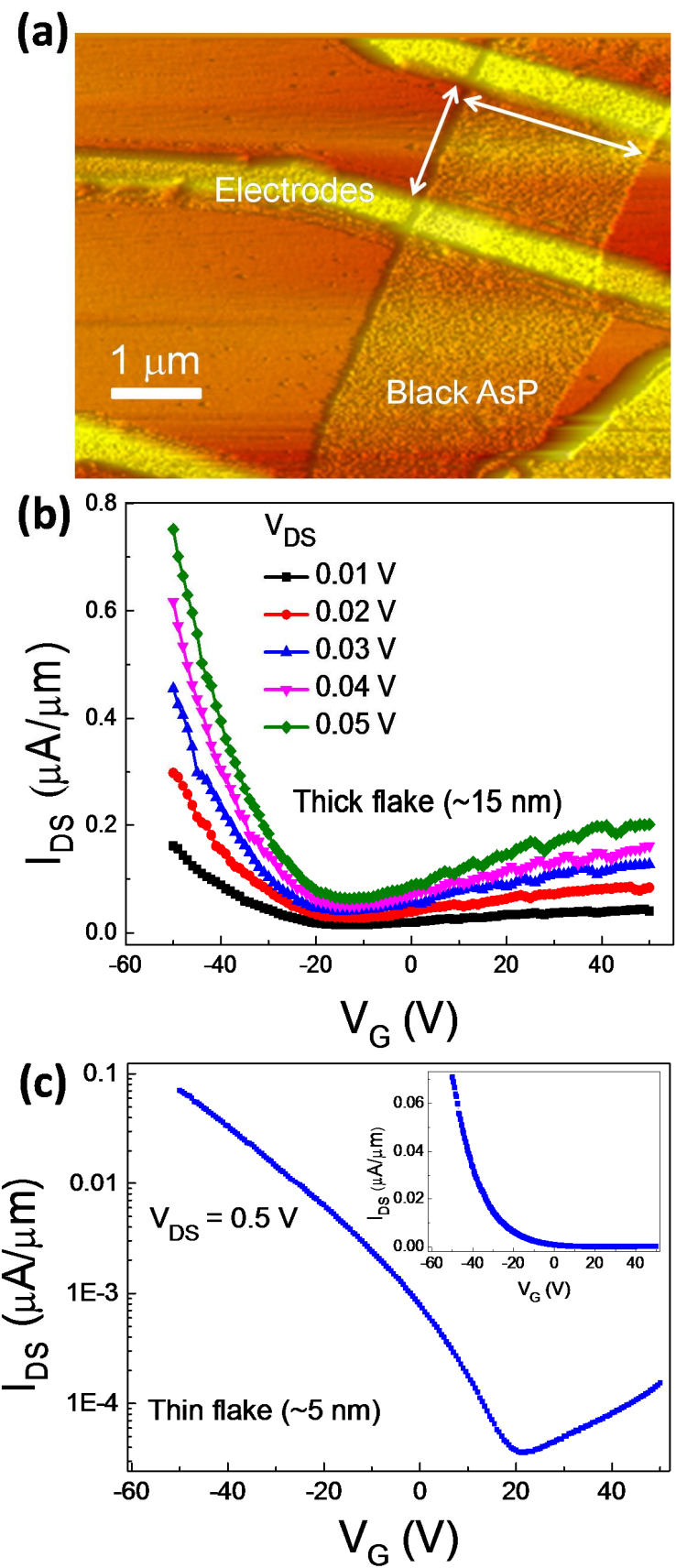

**Figure 2**

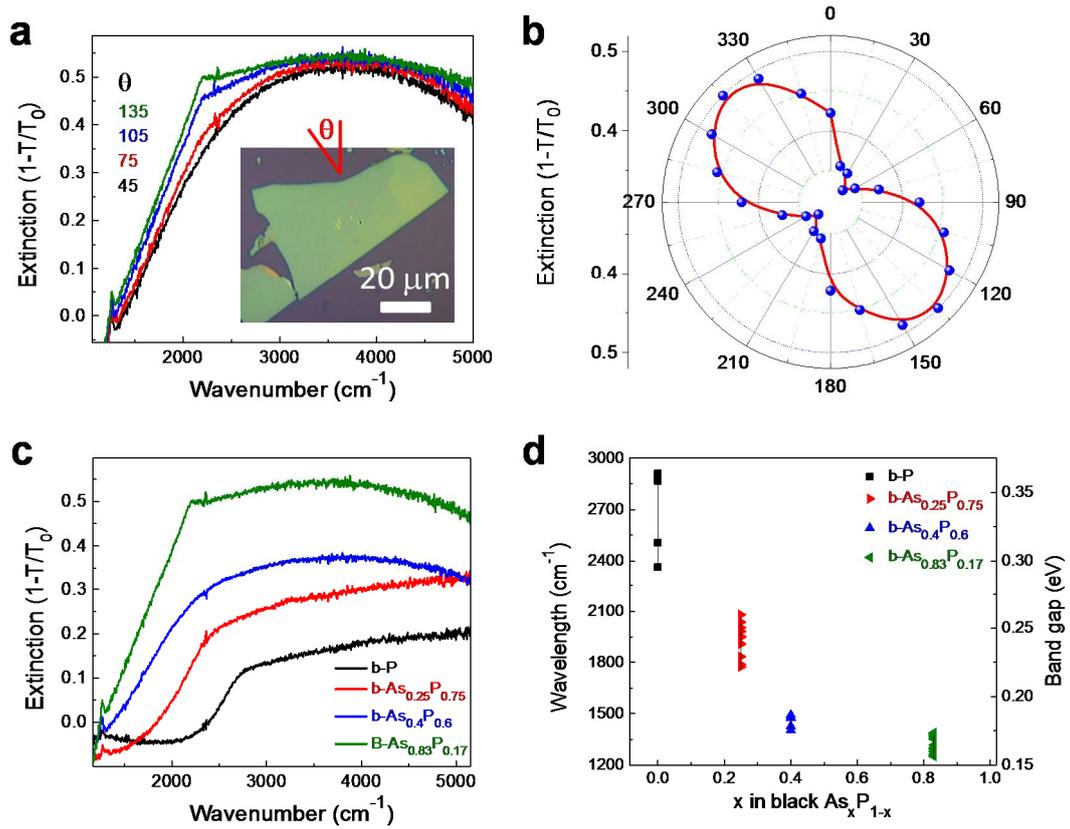

**Figure 3**

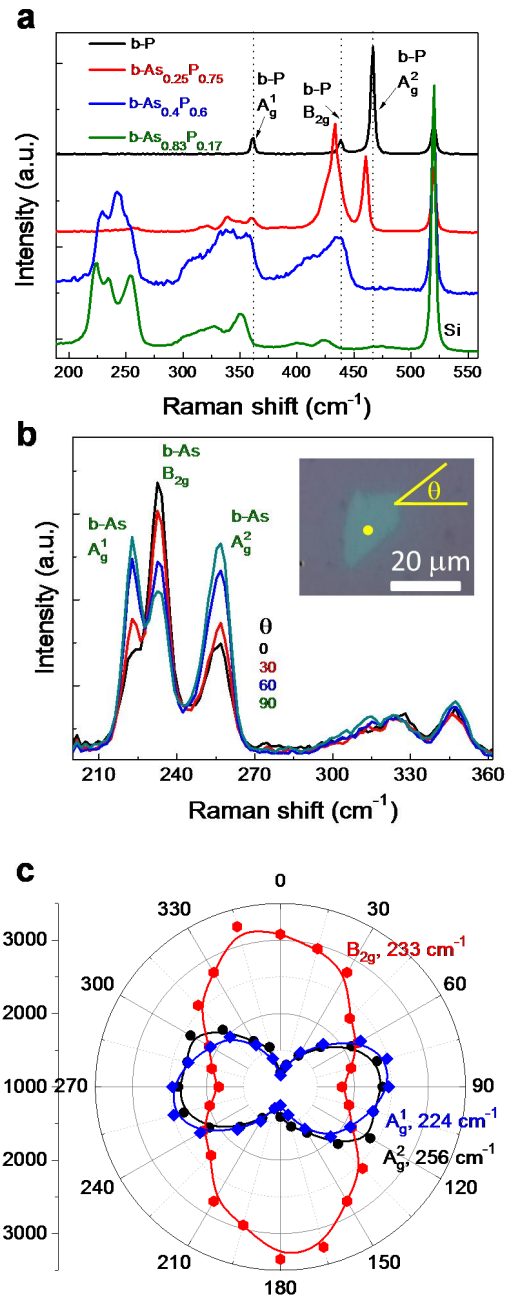

**Figure 4**

**TOC Figure**

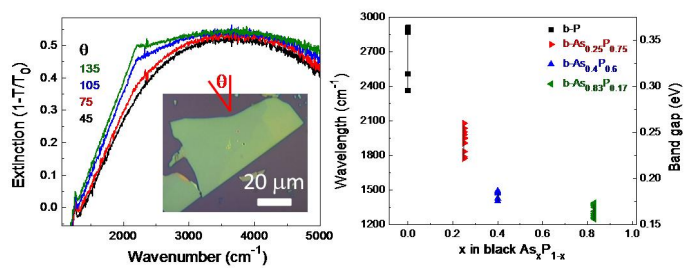

*Short introduction:*

**We introduce new layered anisotropic infrared semiconductors, black arsenic-phosphorus (b-AsP), with highly tunable chemical compositions and electronic and optical properties. Transport and infrared absorption studies demonstrate the semiconducting nature of b-AsP with tunable bandgaps, ranging from 0.3 to 0.15 eV. These bandgaps fall into long-wavelength infrared (LWIR) regime and cannot be readily reached by other layered materials.**

**Black Arsenic-Phosphorus: Layered Anisotropic Infrared Semiconductors with Highly Tunable Compositions and Properties**

*Bilu Liu1, Marianne Köpf2, Ahmad A. Abbas1, Xiaomu Wang3, Qiushi Guo3, Yichen Jia3, Fengnian Xia3, Richard Weihrich4, Frederik Bachhuber4, Florian Pielnhofer4, Han Wang1, Rohan Dhall1, Stephen B. Cronin1, Mingyuan Ge1, Xin Fang1, Tom Nilges2, Chongwu Zhou1\**

Dr. B. Liu, A. Abbas, Prof. H. Wang, R. Dhall, Prof. S. B. Cronin, M. Ge, X. Fang, Prof. C. Zhou
Ming Hsieh Department of Electrical Engineering, University of Southern California, Los Angeles, California, 90089, USA
E-mail: chongwuz@usc.edu

M. Köpf , Prof. T. Nilges
Technische Universität München, Department of Chemistry, Lichtenbergstraße 4, Garching b. München 485748, Germany

Dr. X. Wang, Q. Guo, Y. Jia, Prof. F. Xia
Department of Electrical Engineering, Yale University, New Haven, Connecticut 06511, USA

Prof. R. Weihrich, F. Bachhuber, F. Pielnhofer
Institut für Anorganische Chemie, Universität Regensburg, Universitätsstraße 31, Regensburg 93040, Germany

# Supplementary Information

**Black Arsenic-Phosphorus: Layered Anisotropic Infrared Semiconductors with Highly Tunable Compositions and Properties**

Bilu Liu[1], Marianne Köpf[2], Ahmad A. Abbas[1], Xiaomu Wang[3], Qiushi Guo[3], Yichen Jia[3], Fengnian Xia[3], Richard Weihrich[4], Frederik Bachhuber[4], Florian Pielnhofer[4], Han Wang[1], Rohan Dhall[1], Stephen B. Cronin[1], Mingyuan Ge[1], Xin Fang[1], Tom Nilges[2], Chongwu Zhou[1*]

1. Ming Hsieh Department of Electrical Engineering, University of Southern California, Los Angeles, California, 90089, USA

2. Technische Universität München, Department of Chemistry, Lichtenbergstraße 4, Garching b. München 485748, Germany

3. Department of Electrical Engineering, Yale University, New Haven, Connecticut 06511, USA

4. Institut für Anorganische Chemie, Universität Regensburg, Universitätsstraße 31, Regensburg 93040, Germany

E-mail: chongwuz@usc.edu

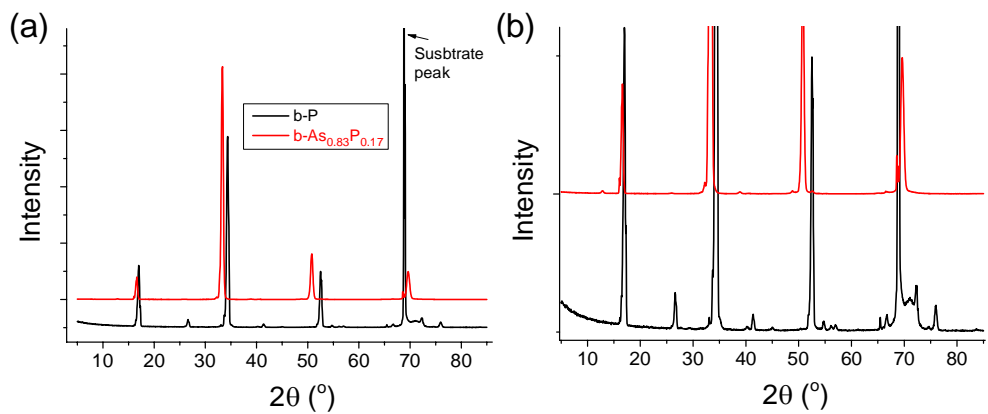

**Supplementary Figure S1** XRD patterns of b-P and b-$As_{0.83}P_{0.17}$ bulk crystals. Plot b is a zoom-in plot of a. The results show that b-AsP has orthorhombic structure, similar to b-P. This result is consistent with our early publication.[1] The XRD patterns were collected using a Rigaku Ultima IV powder/thin-film diffractometer with Cu Kα line.

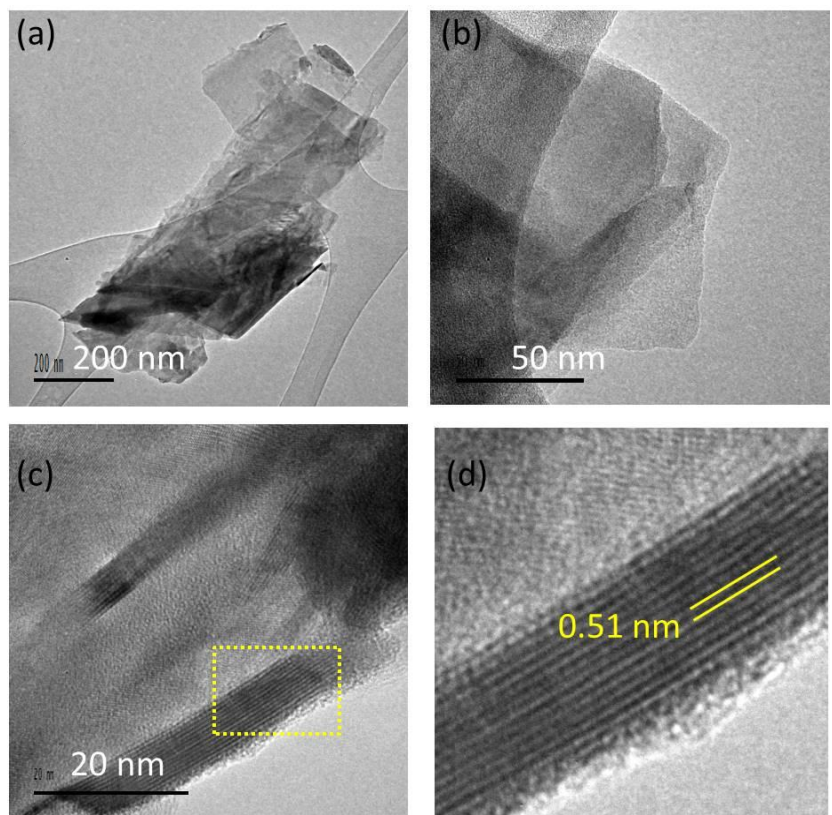

**Supplementary Figure S2** TEM characterization of b-$As_{0.83}P_{0.17}$ samples. (a) (b) Low magnification TEM images show thin flakes of b-$As_{0.83}P_{0.17}$. (c) A HRTEM image shows the edge of a flake. (d) Zoom-in image of the yellow box highlighted area in image c, showing a ~13 layer b-$As_{0.83}P_{0.17}$ flake. The layer distance was measured to be 0.51 nm. For TEM sample preparation, the b-AsP flakes were first exfoliated on scotch tape, followed by gently attached the tape onto a TEM grid. After separation of the tape with TEM grid, some b-AsP flakes left on TEM grid were used for subsequent TEM characterization.

**DFT calculations**

First principles calculations were performed within the framework of DFT with GGA-PBE functionals and Grimme D2 corrections for van-der-Waals interactions, to calculate the geometrical structure, band gap, and vibrational properties of black arsenic-phosphorus (b-AsP) system. This combination provides the best results based on our recent investigations on phosphorus allotropes (see Figure S3 and reference [1]). Here, full structure optimizations and calculations of vibrational frequencies were performed as implemented in the CRYSTAL14 code.[2, 3] For phosphorus and arsenic, all electron basis sets with triple-zeta valence were applied.[4] Additional electronic structure calculations were performed with the full potential local orbital code FPLO14 in the scalar relativistic mode.[5, 6] For the graphical representation of vibrational frequencies, the j-ice was applied.[7]

The calculated band gaps for b-$As_xP_{1-x}$ are listed in Table S1. Systematical underestimated band gaps are due to the GGA functional as known from literature. Nevertheless, the general trend is confirmed as also found from test calculations with the B3LYP and HSE hybrid functionals that overestimate the band gaps. We also find that the calculated electronic structure further depends strongly on distortions in the structure related to the van der Waals interactions. From a shortening of interlayer distances (sample thickness increases), a smaller band gaps are observed, i.e., monolayer b-AsP would exhibit larger band gaps than the bulk. Moreover, with shorter inter-layer distances, the direct band gap is systematically shifted from the Y k-point towards U, due to van der Waals interactions.

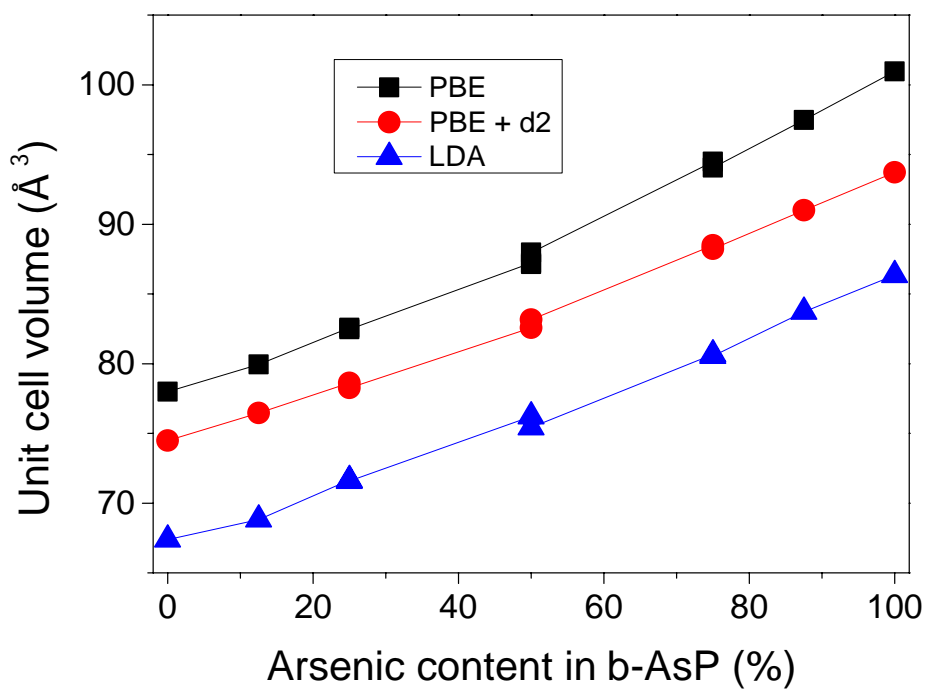

**Supplementary Figure S3** Cell volumes for b-As$_x$P$_{1-x}$ calculated with PBE, PBE+d2, and LDA functionals. The PEB and LDA results are from previous studies[1] and used as comparisons here.

**Supplementary Table S1** Calculated band gaps for b-AsP using GGA method.

| Materials (b-As$_x$P$_{1-x}$) | b-As$_0$P$_1$ (pure b-P) | b-As$_{0.25}$P$_{0.75}$ | b-As$_{0.5}$P$_{0.5}$ | b-As$_{0.75}$P$_{0.25}$ | b-As$_1$P$_0$ (pure b-As) |
|---|---|---|---|---|---|
| Band gaps (eV) | 0.16 | 0.13 | 0.11 | 0.052 | 0.05 |

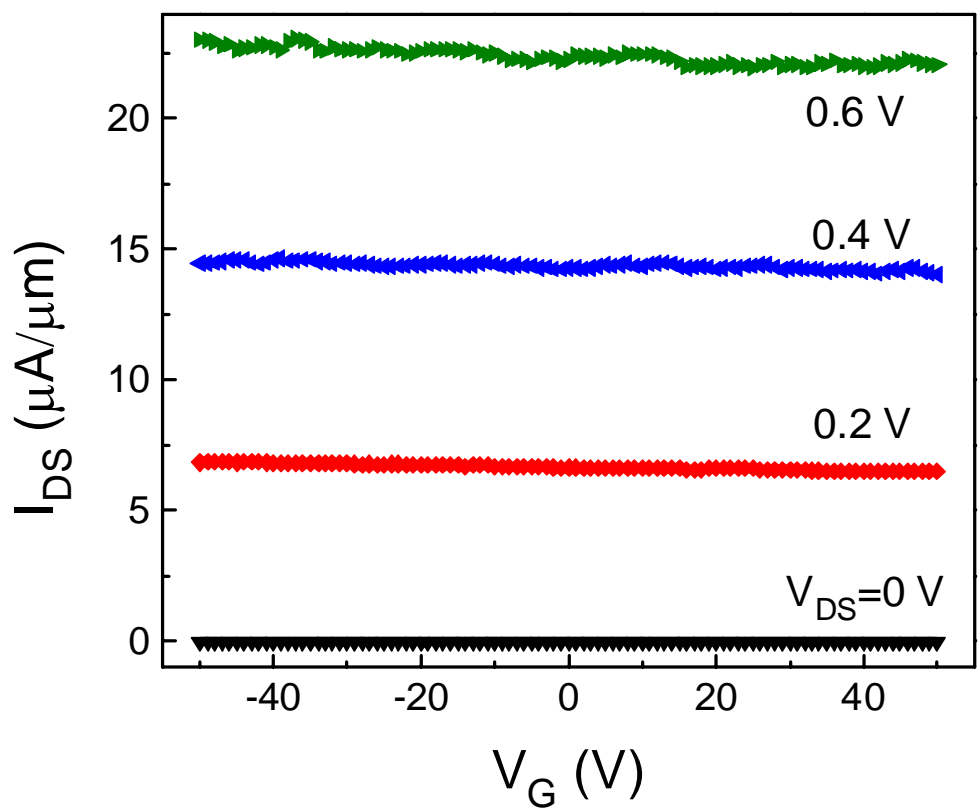

**Supplementary Figure S4** $I_{DS}$-$V_G$ characteristics of a representative thick black arsenic-phosphorus flake (b-As$_{0.83}$P$_{0.17}$). The thickness of the flake is around 60 nm based on AFM measurements. The sample is highly conductive and there is nearly no gate dependence for such thick flakes due to screening effect.

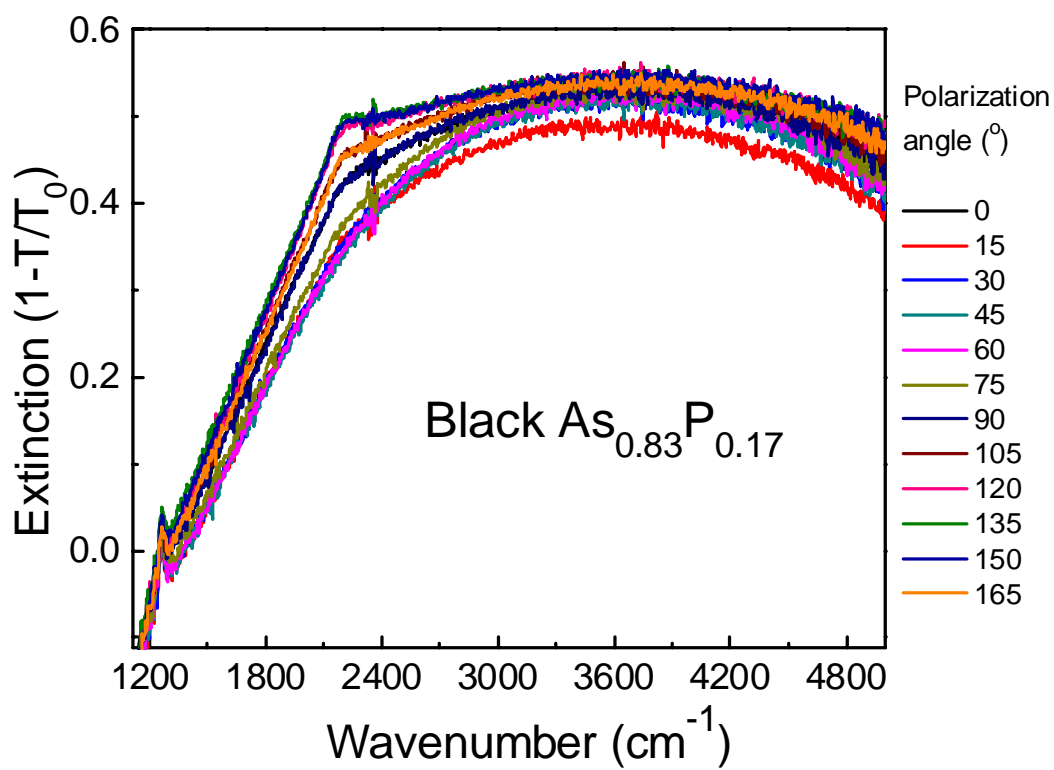

**Supplementary Figure S5** Polarization-resolved infrared absorption spectra of a black-$As_{0.83}P_{0.17}$ flake at different polarization angles of 0 to 165 degrees with a step of 15 degree. Figure 3a in the main text uses part of the spectra (45, 75, 105, and 135 degrees) shown here.

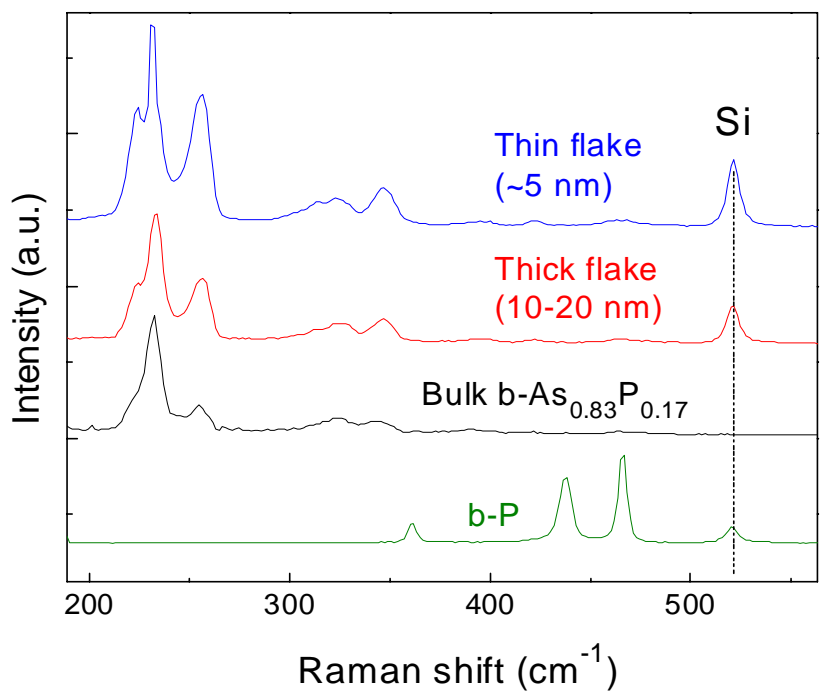

**Supplementary Figure S6** Raman spectra of bulk (black), thick flake (red), and thin flake (blue) of b-As$_{0.83}$P$_{0.17}$. The Raman spectrum of pure b-P (i.e., b-As$_0$P$_1$) is also shown as green curve for comparison. The spectra were vertically shifted for clarify.

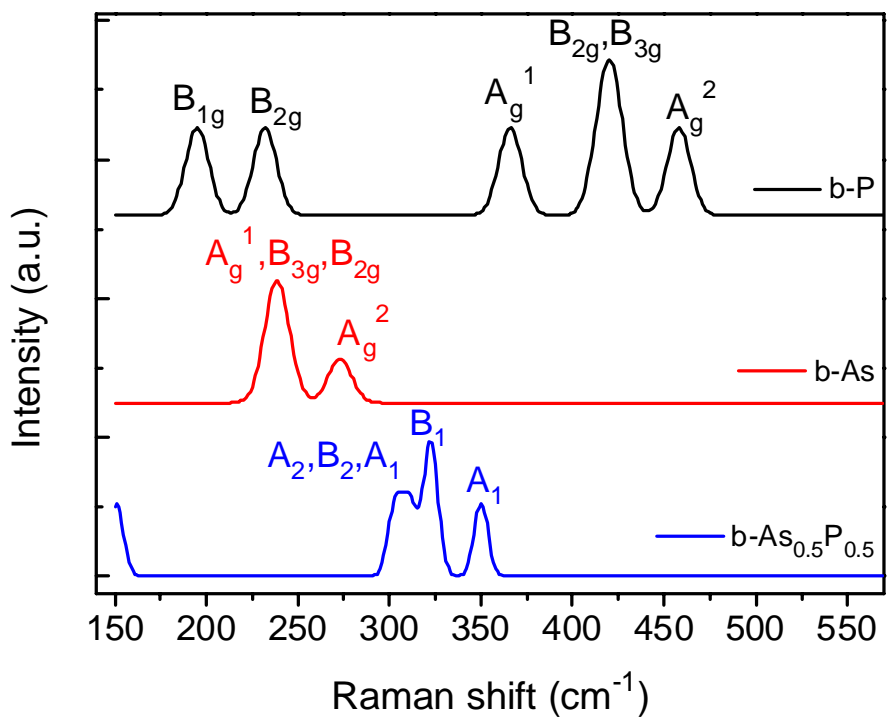

**Supplementary Figure S7** Calculated Raman spectra of black phosphorus (b-P, black), black arsenic (b-As, red), and black arsenic-phosphorus (b-As$_{0.5}$P$_{0.5}$, blue, model contains arsenic-phosphorus bonds only). The calculation method is PBE-d2.

***Supplementary Table S2*** Calculated Raman shift for b-P, b-As, and b-AsP

| Modes | $A_g^2$ (cm$^{-1}$) | $B_{3g}$ | $B_{2g}$ | $A_g^1$ | $B_{2g}$ | $B_{1g}$ |
|---|---|---|---|---|---|---|
| b-P | 458 | 422 | 417 | 365 | 232 | 195 |
| | $A_g^2$ | $B_{2g}$ | $A_g^1$ | $B_{3g}$ | $B_{2g}$ | $B_{1g}$ |
| b-As | 273 | 241 | 237 | 237 | 121 | 103 |
| | $B_1$ | $A_1$ | $B_1$ | $A_2$ | $B_2$ | $A_1$ |
| b-As$_{0.5}$P$_{0.5}$ | 372 | 350 | 324 | 320 | 310 | 302 |

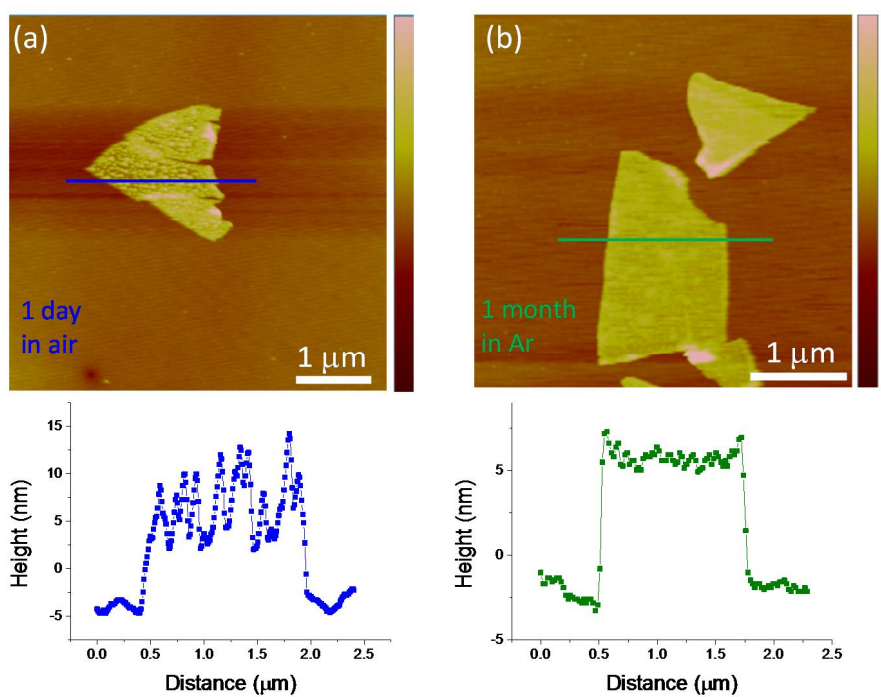

**Supplementary Figure S8** Stability of b-As$_{0.83}$P$_{0.17}$ samples. (a) AFM image of a flake after air exposure for 1 day. (b) AFM image of another flake after stored in Ar glove box for about 1 month. The vertical bars are 40 nm for both images. The flake thicknesses are ~10 nm for both flakes. The bottom plots show height profile along the blue line in image a and green line in image b. The results show that b-As$_{0.83}$P$_{0.17}$ flakes underwent severe degradation in air, while remained stable in Ar atmosphere.

## References

[1] O. Osters, T. Nilges, F. Bachhuber, F. Pielnhofer, R. Weihrich, M. Schoneich, P. Schmidt, *Angew Chem Int Ed* **2012,** *51*, 2994.

[2] J. P. Perdew, K. Burke, M. Ernzerhof, *Phys. Rev. Lett.* **1996,** *77*, 3865.

[3] S. Grimme, *J Chem Phys* **2006,** *124*.

[4] M. F. Peintinger, D. V. Oliveira, T. Bredow, *J Comput Chem* **2013,** *34*, 451.

[5] I. Opahle, K. Koepernik, H. Eschrig, *Phys Rev B* **1999,** *60*, 14035.

[6] K. Koepernik, H. Eschrig, *Phys Rev B* **1999,** *59*, 1743.

[7] P. Canepa, R. M. Hanson, P. Ugliengo, M. Alfredsson, *Journal of Applied Crystallography* **2010,** *44*, 225.